\documentclass[aps, showpacs,prd,superscriptaddress,eqsecnum] {revtex4}
\usepackage{hyperref}
\usepackage{amsmath }
\usepackage{amssymb}
\usepackage{epsfig}
\usepackage{natbib}
\newcommand*{\be}{\begin{equation}}
\newcommand*{\ee}{\end{equation}}

\begin{document}
\bibliographystyle{revtex}
\title{Accelerating Universe and Pioneer anomaly as manifestation of
conformal time inhomogeneity}

\author{L.M. Tomilchik}
\email{lmt@dragon.bas-net.by}

\affiliation{B.I. Stepanov Institute of Physics, 68 Nezalezhnasci
Avenue, 220072 Minsk, Belarus}

\begin{abstract}
The description of the cosmological expansion and its possible
local manifestations via treating the proper conformal
transformations as a coordinate transformation from a comoving
Lorentz reference frame (RF) to an uniformly accelerated RF is
given. The explicit form of the conformal deformation of time is
established. The  expression defining the location cosmological
distance in the form of simple function on the red shift is
obtained. By coupling it with the well known relativistic formula
defining the relative velocity of the mutually moving apart source
and receiver of the signal, the explicit analytic expression for
the Hubble law is obtained. The connection between acceleration
and the Hubble constant follows therefrom immediately. The
expression for the conformal time deformation in the small time
limit leads to the quadratic time nonlinearity. Being applied to
describe the location-type experiments, this predicts the
existence of the uniformly changing blue-shifted frequency drift.
Phenomenon of the Pioneer Anomaly (PA) is treated as the first of
such a kind of effects discovered experimentally. The obtained
formulae reproduce the PA experimental data. The expression
generalizing the conventional  Hubble law reproduces the
experimentally observed phenomenon which in the frame of the
conventional cosmological paradigm is treated as the transition
from the decelerated expansion of the Universe to the accelerated
one.
\end{abstract}


\maketitle

\section{Introduction}

The experimental detection of the Pioneer Anomaly (PA) [1--4]
indicates on the existence, in the outer of the solar system at
the distances \~20--70 AU, an observable effect unaccountable via
contemporary celestial mechanics including all the necessary
General Relativity (GR) corrections.

The experimentally observed phenomenon is the anomalous blue
frequency drift with a numerical value $\frac{d\nu}{dt}=(5,99\pm
0,01)10^{-9}$ Hz/s detected in the signals retransmitted by
Pioneer 10 and 11 satellites. It is interpreted as the Doppler
shift, that is, due to the uniform acceleration directed (within
the limits of experimental error) towards the Sun amounting, after
all corrections, to $a_p=(8,74\pm 1,33)10^{-8}$ cm/s$^2$ (see
[1--4]).

The fact that the observable $a_p$ is close to the quantity
$W_0=cH_0$ ($c$ is the speed of light, $H_0$ is the Hubble
constant) and therefore can point to the cosmological origin of
the PA effect, was noted by several authors (see, for example,
[2], sec. XI, point C). However, any unambiguous theoretical
arguments in favor of the possible existence of such a connection
are absent up to now.

Moreover, the rigorous consideration of this subject on the basis
of the precise Friedmann and Schwarzschild solutions of the GR
equations gives the quadratic dependence of the effect on the
parameter $H_0$ [5], so that the predicted magnitude of $a_p$
turns out to be much less than the observable one and, in
addition, the effect was found to have the opposite sign [2].
Nevertheless, an interpretation of PA as a local manifestation of
the cosmological expansion seems to be acceptable when using the
possibilities inherent in the general symmetry group of the
Minkowski space.

Most recent experimental data of astrophysics clearly indicate an
infinitesimal space curvature of the observable Universe. Because
of this, it seems natural to select the conformally Minkowski
space as a model for space-time manifold of the modern Metagalaxy.
As is well known, its most wide symmetry group is the group
$SO(4.2)$ which, apart from the Poincar\'{e} group isometric
transformations, includes a subgroup of special conformal
transformations (SCT) and dilatations changing the space and time
scales [6]. Naturally, it may be expected that an expansion of the
space-time symmetry by the inclusion of this subgroup
transformations will result in some generalization of the standard
kinematics of the Special Relativity (SR).

Such an attempt was undertaken by us before. In the paper [7], on
the basis of the nonisometric transformations subgroup of the
$SO(4.2)$ group, the nonlinear time inhomogeneity one-parameter
conformal transformations were constructed. After postulating the
connection between the group parameter and the Hubble constant it
is shown that the existence of an anomalous blue-shifted frequency
drift is the pure kinematic manifestation of the time
inhomogeneity induced by the  Universe expansion. The obtained
formulae reproduce the observable Pioneer Anomaly effect.
According to the proposed approach, the anomalous blue-shifted
drift is universal, does not depend on the presence of gravitating
centers and can be, in principle, observed at any frequencies
under suitable experimental conditions. The explicit analytic
expression for the speed of recession~--- intergalactic distance
ratio is obtained in the form of a simple function of the red
shift $z$, valid in the whole range of its variation. In the small
$z$ limit this expression exactly reproduces the Hubble law. The
existence of maximum of this function at $z=0.475$ quantitatively
corresponds to the experimentally found value
$z_{exp}=0.46\pm0.13$ of the transition from the decelerated to
the accelerated expansion of the Universe.

As it is known, the special conformal transformations (SCT) can be
treated as a coordinate transformation from an inertial reference
frame (RF) to the uniformly accelerated one. Such a consideration
(the ''new relativity'') was proposed in due time in the papers
[8], and was subsequently discussed by a number of authors (see
[9, 10] and references therein). But in the paper [7] this
interpretation was not employed in the explicit and successive
form.

In the present paper the description of the cosmological expansion
and its possible local manifestations is given via treating the
proper conformal transformations as a coordinate transformation
from a comoving Lorentz RF to an uniformly accelerated RF. Such an
approach permit us to derive the explicit analytic expression for
the Hubble law, which allows us to connect the acceleration
parameter with the Hubble constant as well as to reproduce the
accelerating Universe effect in exact correspondence with the
observations.

In so doing these effects, once dictated by the conformal time
inhomogeneity, can be interpreted as the observable manifestations
of the background acceleration existence, i.e., manifestations of
the noninertial character of any physical frame of reference which
is locally coupled to an arbitrary point of the modern expanding
Metagalaxy.

The content of the present paper is as follows. In section 2 we
consider the proper conformal transformations in the
two-dimensional subspace of the Minkowski space treated as a
coordinate transformations from the local comoving Lorenz
reference frame to the noninertial (uniformly accelerated) one
including the transfer to Galilei-Newton (GN) kinematic limit.

The general expression for the velocity transformation is obtained
in section 3. In the GN limit, the relation brings into existence
the blue Doppler shift which is to be observed as a manifestation
of the noninertiality of the observer reference frame.

In section 4, the general formula for the acceleration
transformation is obtained. It is shown that the parameter of the
conformal transformations defines the background acceleration
which is to be observed in the absence of any real sources of the
dynamical subjection.

In section 5, it is shown that SCT, which is preserving the light
cone equation, nevertheless brings into existence the nonlinear
transformation of its generating lines (conformal deformation of
time or the time inhomogeneity). We establish the explicit form of
this transformation. The parameter having the dimension of time
(the limiting time $t_{lim}$) is connected with the acceleration
parameter $w$ ($t_{lim}=c/w$, $c$ is the speed of light).

Then in section 6, the explicit expression defining the location
cosmological distance $R(z)$ in the form of simple function on the
red shift $z$ is obtained as a direct outcome of the formula
defining the conformal time deformation. Coupling it with the well
known relativistic formula defining the relative velocity $V(z)$
of the mutually moving apart source and receiver of the signal, we
obtain the explicit analytic expression for the ratio
$\phi(z)=V(z)/R(z)$. The function $\phi(z)$ reproduces, in the
limit $z\rightarrow0$, the conventional form of the Hubble law. It
follows herefrom immediately the connection between acceleration
parameter $w$ and the Hubble constant $H_0$: $w=\frac12cH_0$.

The expression describing the conformal time deformation in the
small time limit ($t/t_{lim}\ll1$) leads to the quadratic time
nonlinearity. Such a time nonuniformity, being applied to describe
the location-type experiments, predicts the existence of the
uniformly changing blue-shifted frequency drift. Its magnitude
$\Delta\nu_{obs}$ is directly connected with the Hubble constant
according to $\Delta\nu_{obs}=\nu_0H_0t$ (where $\nu_0$ is the
frequency of the signal emitted, $t$ is the time of the signal
propagation from the emitter to the receiver). Phenomenon of the
PA is treated as the first of such a kind of effects discovered
experimentally. The obtained formulae reproduce the PA
experimental data (section 7).

The formula for the function $\phi(z)=V(z)/R(z)$ generalizing the
conventional expression for the Hubble law is derived by a pure
kinematical considerations. But, owing to the availability of a
maximum of this function at the point $z_0\simeq0.475$, this
formula, in fact, reproduces the experimentally observed
phenomenon which, in the frame of the conventional cosmological
paradigm, is treated as the transition from the decelerated
expansion of the Universe to the accelerated one.

\section{Conformal transformations and the accelerated frame of reference}

It is well known that the special conformal transformations (SCT)
$$x'^\mu = \frac{x^\mu +a^\mu(x^\alpha
x_\alpha)}{1+2(a^\alpha x_\alpha)+(a^\alpha a_\alpha) (x^\beta
x_\beta)}, \eqno(1)$$ where $a^\mu$ is the four-vector parameter,
$\eta_{\mu\nu}=\mathrm{diag}(1,-1,-1,-1)$, can be interpreted as
the transformations between the Lorentz (inertial) frame of
reference $S(x^\mu)$ and the noninertial (accelerated) frame of
reference $S'(x'^\mu)$ (see, for example, [10]). Following [10] we
shall for simplicity consider a two-dimensional subspace
$\{t,x\}$, i.e. we put $$x^\mu=(ct,x,0,0),\
a^\mu=(0,-\frac{w}{2c^2},0,0),\eqno{(2)}$$ where $w$ is a
constant, having the dimension of acceleration.

Let us write the transformations (1) in the following noncovariant
form: $$\left.\begin{gathered}x'=
\frac{x\left(1+\frac{wx}{2c^2}\right)-\frac{wt^2}{2}}
{\left(1+\frac{wx}{2c^2}\right)^2-\left(\frac{wt}{2c}\right)^2},\\
t'=\frac{t}{\left(1+\frac{wx}{2c^2}\right)^2-\left(\frac{wt}{2c}\right)^2}.
\end{gathered}\right\} \eqno{(3)}$$

In the case when $\frac{wx}{2c^2}$ and $\frac{wt}{2c}$ are
negligible from (3) we have $$x'=x-\frac{wt^2}{2},\qquad
t'=t,\eqno{(4)}$$ which corresponds to Galilei-Newton kinematics.
It should be noted that the identification of $w$ with a constant
three-dimensional newtonian acceleration is essentially based on
this correspondence. It is also clear from (3) that sign $(-)$ of
the vector's $a^\mu$ $x$-component describes a positive direction
of $S'$ acceleration along the $x$-axis of the inertial reference
frame (IRF) $S$.

The following note deals with the identification of the RF $S$. It
is well known that the parameter $a^\mu$ is related with a
constant 4-acceleration $W^\mu$ as follows:
$$a^\mu=\frac1{2c^2}W^\mu.$$ From the other side, generally, the
4-vector of acceleration is
$$W^\mu=(W^0,\vec W)=\gamma^2\left(\gamma^2\frac{(\vec v\vec w)}{c},
\frac{(\vec v\vec w)}{c^2}\vec v+\vec w\right),\eqno{(5)}$$ where
$\displaystyle\vec v=\frac{d\vec r}{dt},\ \vec w=\frac{d\vec
v}{dt},\ \gamma=\left(1-\frac{\vec v^2}{c^2}\right)^{-\frac12}$.

From relation (5) it follows that in the local comoving reference
frame (CRF), i.e., in the frame where $\vec v=0$, time component
of the 4-acceleration is exactly null, and its space part is $\vec
W=\vec w$: $$W^\mu=(0,\vec w).$$ Because of this, the parameter
$a^\mu$ of the form (2) determines the 4-acceleration in CRF.
Therefore, the 4-vector $x^\mu$ in the transformations (1) is
supposed to be determined in the CRF.

We also emphasize that, due to (3), the transformations (1) are
singular when $$\left(1+\frac{wx}{2c^2}\right)^2=
\left(\frac{wt}{2c}\right)^2,\eqno{(6)}$$ i.e. on the lines
$$x_\pm=-\frac{2c^2}{w}\pm ct,\eqno{(7)}$$ corresponding to the null
cone of the $(0,-2r_0)$ point (see Fig. 1).

\begin{figure}[h!]
\centering \includegraphics[scale=0.7]{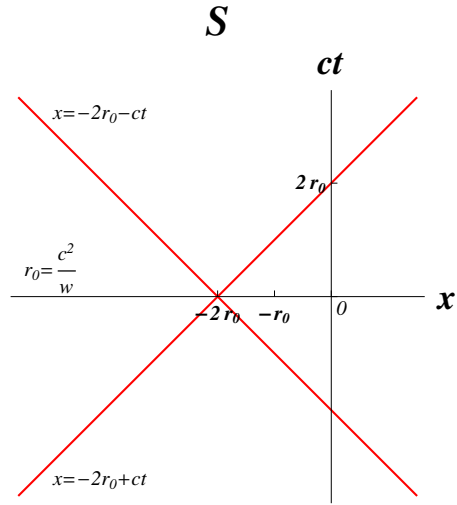} 
\caption{Lines of transformations (3) singularity: a light cone
with center at $(0,-2r_0)$. Every fixed acceleration parameter $w$
defines an event horizon (Rindler horizon) for the observer at
$x'=0$ in RF $S'$.}
\end{figure}

It is well known that under the transformations (3) the world
lines $\{x_0,x_{fix}\}$ in the $(x_0,x)$ plane, where $x_{fix}$ is
fixed, convert to hyperbolas in the $(x'_0,x')$ plane. Again
following [10], we denote the world lines $\{x_0,x_{fix}\}$ as
$\{x_0,\alpha\}$, where $\alpha=1+\frac{x_{fix}}{2r_0}$, such that
$x_{fix}$ is defined as a function of non-zero real parameter
$\alpha$: $$x=2r_0(\alpha-1), \qquad
(\alpha\gtrless0).\eqno{(8)}$$ Then the world lines
$\{x_0,x_{fix}\}$ transform into hyperbolas
$$\left(x'-\lambda(\alpha)\right)^2-{x'}^2_0=
\left(\frac{r_0}{\alpha}\right)^2\eqno{(9)}$$ in the $(x'_0,x')$
plane. Here $$\lambda(\alpha)=\frac{r_0}{\alpha}(1-2\alpha)
\eqno{(10)}$$ defines the hyperbola center position
$(x'_0=0,x'=-\lambda)$. Vertex of hyperbola is determined from (9)
with $x'_0=0$ $$x'_\pm(\alpha)=-\frac{r_0}{\alpha}+2r_0\pm
\frac{r_0}{\alpha}=\left\{\begin{aligned}&2r_0,\\
&2r_0\frac{\alpha-1}{\alpha}.\end{aligned}\right.\eqno{(11)}$$ Two
such lines are showed on Fig. 2.

\begin{figure}[h!]
\centering \includegraphics[width=0.45\textwidth]{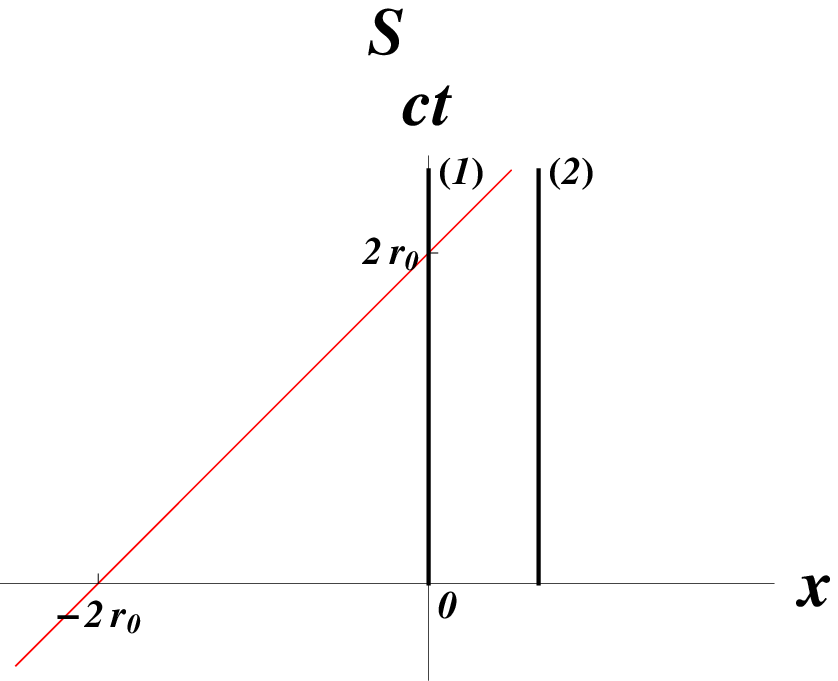}
\includegraphics[width=0.48\textwidth]{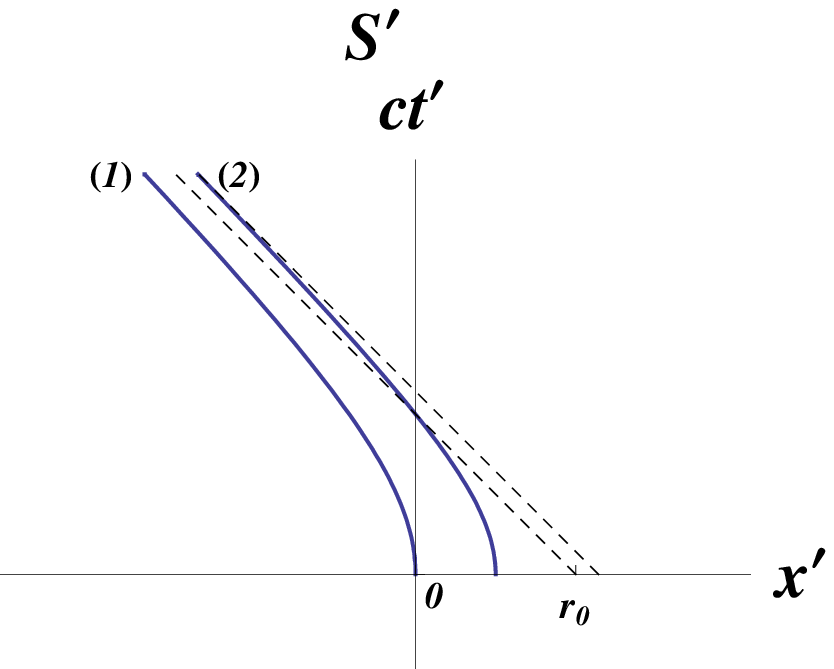} 
\caption{World lines of fixed particles (1) and (2) transforms
into hyperbolas with $\alpha=1$ and $\frac43$, respectively from
left to right. Asymptotes of the hyperbolas are the dashed lines.
For the sake of simplicity only positive time values are shown.}
\end{figure}

The meaning of the transformation is clear: every point fixed in
Lorentz RF $S$ is moving hyperbolically in non-inertial RF $S'$.

Let us now consider the velocity and acceleration transformation
laws.

\section{Velocity transformation and the blue-shifted Doppler drift}

The transformation of coordinate differentials is convenient to
express by (3) in the following symmetric form:
$$\left.\begin{aligned}
dx'= \frac{(\xi^2+\eta^2)dx-2\xi\eta dx_0}{\xi^2-\eta^2},\\
dx'_0= \frac{(\xi^2+\eta^2)dx_0-2\xi\eta dx}{\xi^2-\eta^2}.
\end{aligned}\right\}\eqno{(12)}$$
Here $$\xi=1+\frac{x}{2r_0},\quad \eta=\frac{x}{2r_0},\quad
x_0=ct, \quad x'_0=ct', \quad r_0=\frac{c^2}{w}.\eqno{(13)}$$ This
results in the following expression for velocity transformations
$\displaystyle V'_x=\frac{dx'}{dt'},\ V_x=\frac{dx}{dt}$:
$$V'_x=\frac{V_x-V_w}{1-\frac{V_xV_w}{c^2}},\eqno{(14)}$$
where the ''effective relative velocity'' of $S$ and $S'$ RF $V_w$
is $$V_w=c\frac{2\xi\eta}{\xi^2+\eta^2}\leqslant c.\eqno{(15)}$$
The value of $V_w$ is a function of space and time coordinates. It
is notable that form of (14) is exactly the same as of usual
velocity transformations in Special Relativity. In the
non-relativistic approximation $\xi\approx1,\ \eta\ll1$ we obtain
$$V'_x=V_x-wt,\eqno{(16)}$$ corresponding to Galilei-Newton
kinematics. The maximum of $V_w$ is determined by $|\xi|=|\eta|$,
and is equal to $c$.

From (16), the existence of blue Doppler drift immediately
follows. Really, due to (16) the longitudinal component of a point
velocity $V'_x$ measured by an observer fixed in non-inertial RF
$S'$ will be less then that measured by an observer fixed in the
inertial comoving RF $S$ by $\Delta V_x=wt$.

Clearly, in this case an observer with fixed space coordinates in
RF $S'$ measures blue frequency shift as compared to an observer
fixed in $S$. In fact, measured frequency of the signal of a
moving away emitter in the non-relativistic limit will be
$$\begin{aligned}&\nu'=\nu_0\left(1-\frac{V'_x}{c}\right)
&\mbox{in RF $S'$,}\\
&\nu'=\nu_0\left(1-\frac{V_x}{c}\right)=\nu_{mod} &\mbox{in RF
$S$,}\end{aligned}$$ where $\nu_0$ is the signal frequency emitted
by a source fixed in $S$.

Therefore, for the observed blue shift we have
$$\Delta\nu_{obs}=\nu'-\nu_{mod}=\nu_0\frac{wt}{c},\eqno{(17)},$$
and $\nu_{mod}$ is the frequency defined by neglecting
non-inertiality of $S'$. In the approximation considered the shift
is linear in time. The rate of shift $\dot\nu_{obs}=
\frac{d\nu_{obs}}{dt}$ is defined by the following simple relation
$$\dot\nu_{obs}=\nu_0\frac{w}{c}.\eqno{(18)}$$

This result, in principle, is well known. Here we have to do, in
fact, with the effect of the gravitational (Einstein) frequency
drift described on the basis of the equivalence principle. The
shift is blue because the non-inertial observer and the source of
signal approach each other, and therefore the magnitude of the
equivalent gravitational potential at the observation point is
above its magnitude at the point of the signal emission.

\section{Transformation of accelerations. The background acceleration}

The transformation law of the acceleration's $x$-component
$W'_x=\frac{d^2x'}{{dt'}^2}$ can be derived directly from (14).
Elementary calculus give
$$W'_x=\frac{(\xi^2-\eta^2)^4}{(\xi^2+\eta^2)^3}
\left(1-\frac{V_wV_x}{c^2}\right)^{-3}\left\{W_x-
\frac{c^2}{r_0}\left(1-\frac{V_x^2}{c^2}\right)
\left(1+\frac{1}{2r_0^2}(x-V_x t)\right)\right\}.\eqno{(19)}$$

In the comoving RF ($V_x=0$), the above formula becomes
$$W'_x=\frac{(\xi^2-\eta^2)^4}{(\xi^2+\eta^2)^3}(W_x-\xi w).$$
In the approximation of $\xi\approx1,\ \eta\ll1$ we have therefrom
$$W'_x=W_x-w.\eqno{(20)}$$ This relation holds in every comoving
RF as long as in this frame $\xi\approx1,\ \eta\ll1$ approximation
is valid. A probe particle free of dynamical influence in the
comoving RF, i.e. with $W_x=0$, in accordance with (16) will be
uniformly accelerated in the non-inertial RF $S'$ with the
constant acceleration of $-w$. Such an acceleration can be
registered by any observer fixed in any point of this non-inertial
RF $S'$. By the equivalence principle the non-inertial observer is
entitled to identify this acceleration with an existence of a
constant (background) gravitational field which results in
acceleration $w$.

\section{Conformal deformation of the light cone and time inhomogeneity}

Now we consider the transformations of the light cone generatrices
under SCT (1). Since $x'^\mu x'_\mu=x^\mu x_\mu(1+2(ax) +
(a)^2(x)^2) ^{-1}$, transformations (1) leave the light cone
equation invariant, i.e., from $x^\mu x_\mu=0$ follows $x'^\mu
x'_\mu=0$. However the light cone surface is deformed
non-linearly. From (1), it generally follows
$$x'^\mu=\frac{x^\mu}{1+2(ax)},\eqno{(21)}$$ when additionally
$x^\mu x_\mu=x'^\mu x'_\mu=0$.

In the two-dimensional case considered we have the relation
$$t'_\pm=\frac{t}{1\pm\frac{t}{t_{lim}}},\eqno{(22)}$$ where
$t_{lim}=c/w$.

The choice of sign corresponds to signal propagation in the
forward and backward directions, respectively. We are reminded
that the symbol $t$ ($t'$) represents time of a light signal
propagation between two spatially separated points in the space of
Lorentz RF $S$ (in the non-inertial RF $S'$). So the quantities
$R=ct$ and $R'=ct'$ define location distances in both of these
RFs.

Obviously a semi-infinite time interval $0\leqslant t<\infty$
corresponding to the positive (forward) direction of signal
propagation maps onto a finite time interval $0\leqslant t'_+
\leqslant t_{lim}$. For the backward direction, on the contrary, a
finite interval $0\leqslant t \leqslant t_{lim}$ maps onto a
semi-infinite time interval $0\leqslant t'_-<\infty$ (Fig. 3).

\begin{figure}[h!]
\centering \includegraphics[width=0.48\textwidth]{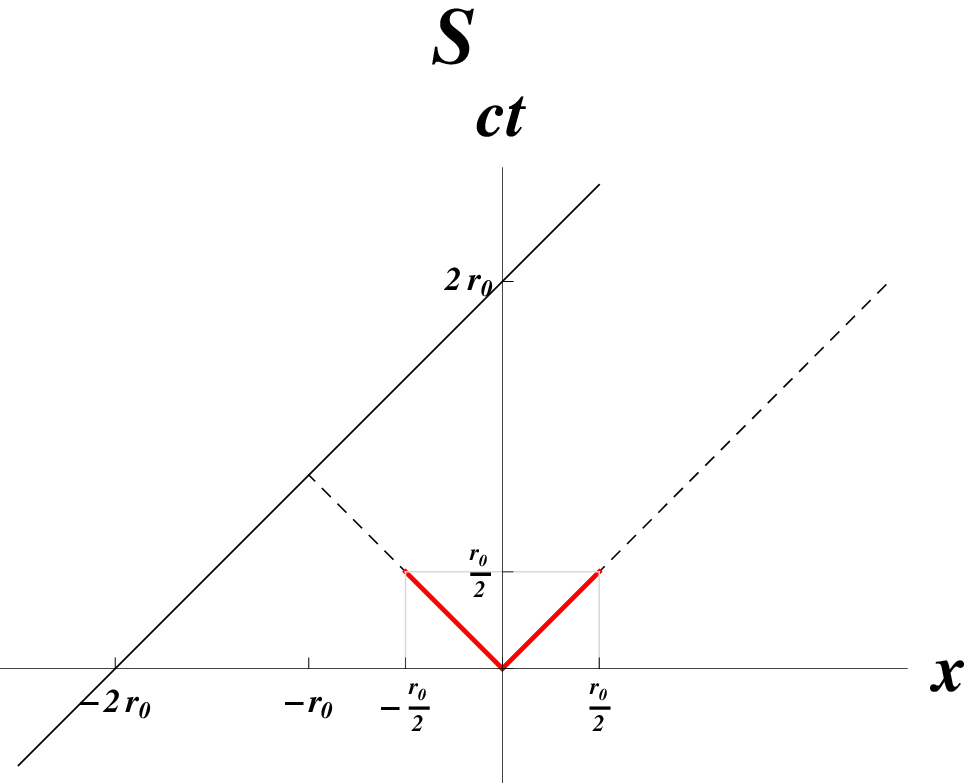}
\includegraphics[width=0.48\textwidth]{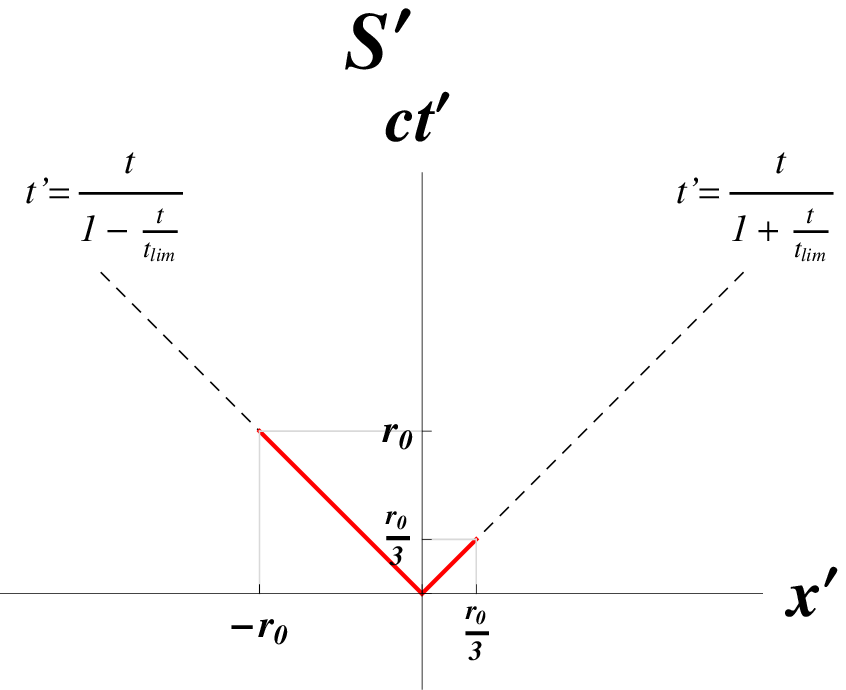} 
\caption{Conformal deformation of the light cone. Initially equal
intervals $t_\pm=\frac{t_{lim}}{2}$ of forward and backward
generatrices transform into unequal intervals
$t'_+=\frac{t_{lim}}{3}$ and $t'_-=t_{lim}$.}
\end{figure}

The non-linear time transformation (22) will further be referred
to as the conformal deformation of time, or conformal time
inhomogeneity.

\section{The dependence of distance on red shift. The Hubble law}

First and foremost we show that transformations (22) allow us to
obtain an explicit expression for location distance as a simple
function of red shift $z$.

Let us consider, on the basis of the formula (22), the case of a
signal propagation from the deep past to the point of the observer
position. That means that we choose the lower sign in the formula
(22): $$t'=\frac{t}{1-\frac{t}{t_{lim}}},\eqno(23)$$ where
$t_{lim}=c/w$, and $t$ is the time of the signal propagation to
the point of observation.

First of all, from (23) we obtain the explicit expression for the
time interval of the signal propagation as a function of the red
shift $z$.

From (23), we have, for small time increments $\Delta t'$ and
$\Delta t$, the following expression
$$
\Delta t'=\Delta t(1-\frac{t}{t_{lim}})^{-2}. \eqno(24)
$$
If $\Delta t$ and $\Delta t'$ are the periods of oscillations of
the emitted ($\Delta t= T_{emitted}$) and received ($\Delta
t'=T_{observable}$) signals, respectively, then, using the
standard definition of the red shift
$$
\frac{\lambda_{observable}}{\lambda_{emitted}}=z+1, \eqno{(25)}
$$
where  $\lambda_{observable}=cT_{observable}$ and
$\lambda_{emitted}=cT_{emitted}$, we find, from (24), the
expression
$$
\frac{\lambda_{observable}}{\lambda_{emitted}}=
(1-\frac{t}{t_{max}})^{-2}=z+1, \eqno(26)
$$
which gives
$$t(z)=t_{max}\frac{(z+1)^{1/2}-1}{(z+1)^{1/2}}.\eqno(27)$$
Here $t(z)$ represents the time interval between the moments of
emitting and receiving the light (electromagnetic) signal. So,
assuming that the speed of light is constant and does not depend
on the velocity of the emitter, the quantity $R=ct$ can be
regarded as the distance covered by the signal.

From the formula (27), we obtain an expression which determines
the explicit form of dependence of $R$ on the red shift
$z$\footnote{Relation of the type $R(z)=const\cdot z$, that
connects cosmological distance $R$ with the red shift, seems to be
first obtained from the conformal symmetry arguments  by Ingraham
[11] in 1954.}:
$$
R(z)=R_u
\frac{(z+1)^{1/2}-1}{(z+1)^{1/2}}=R_u\left(1-\frac{1}{(z+1)^{1/2}}\right).
\eqno(28)
$$
Here$R_u=ct_{lim}$ is a parameter, which, within the model
suggested, has the sense of the limit (maximal) distance.

The quantity $R(z)$ defined by (28) corresponds to the distance,
which in cosmology is referred to as a location distance. In
principle, the relation (28) allows for a direct experimental
verification in the whole range of $z$ variation, and can be
confirmed or refused by observations.

Now, we can obtain the explicit expression for the Hubble law. For
this purpose, we make use of the well-known formulae describing
the Doppler effect in Special Relativity. In the context of our
approach, to find the explicit expression for the longitudinal
Doppler effect, it is convenient to apply the formulae which
immediately follow from the Lorentz boosts written in the
light-cone variables:
$$
u=\frac{1}{2}(x_0+x),\;\;\;\; v=\frac{1}{2}(x_0-x).
$$
These expressions are (see, for example, [12]):
$$
u'=k(\beta)u,\qquad v={k}^{-1}(\beta)v, \eqno(29)
$$
where $k(\beta)=(\frac{1-\beta}{1+\beta})$,  $\beta=V/c$, $V$ is
the velocity which can be identified with the radial component of
relative velocity of emitter and receiver motion.

Clearly, in terms of $u$ and $v$, the Lorentz boosts have the form
of dilatations. For the description of the Doppler effect, we need
to use the equation of light cone $4uv=0$. Then, for the case of
the signal traveling in positive ($v=0$) and negative ($u=0$),
directions we have in coordinates ($x, t$):
$$
t'=\left(\frac{1\mp\beta}{1\pm\beta}\right)^{1/2} t, \eqno(30)
$$

In the case of emitter and receiver moving away from each other,
we find, from (30), for small time increments $\Delta t'$ and
$\Delta t$:
$$
\Delta t'=\Delta t \left(\frac{1+\beta}{1-\beta}\right)^{1/2},
$$
whence, taking into account (30), the known expression for the
function $V(z)$ follows:
$$
\frac{V(z)}{c}=\frac{(z+1)^2-1}{(z+1)^2+1}. \eqno(31)
$$
Equation (31) is valid in the whole range of the velocity $V$
variation.

We again emphasize the essentially kinematic nature of the
relation (28). It is the manifestation of the
\underline{nonlinear} conformal time deformation (22) which
follows from Special Conformal Transformations exactly in the same
manner as the Doppler effect, and the dependence $V(z)$ follows
from the \underline{linear} time deformation (30) arising from the
Lorentz boosts leaving the equation of light cone unaltered.

Now we can find, using (28) and (31), the following expression for
the ratio $V/R$: $$ \frac{V(z)}{R(z)}=cR_u^{-1}f(z),\eqno{(32)}$$
$$ f(z)=\frac{(z+1)^{1/2}}{(z+1)^2+1}
\cdot\frac{(z+1)^2-1}{(z+1)^{1/2}-1}. \eqno{(33)}$$ It is easy to
see that $\lim_{z\rightarrow0}f(z)=2$.

In this limit we obtain the conventional expression for the Hubble
law:
$$V=H_0R,\eqno(34)$$ where $H_0$ is the Hubble constant.

By comparing this formula with $\lim_{z\rightarrow0}
\frac{V(z)}{R(z)}$ from formula (32), we can establish the
following connection between the acceleration $w$ and the Hubble
constant $H_0$: $$2cR_u^{-1}=2w=cH_0.\eqno(35)$$ It is seen that
the relation (22) defining the conformal deformation of time (the
time inhomogeneity) allows us to establish the following simple
connection between the parameter $w$ defining the background
acceleration and the Hubble constant $H_0$
$$w=\frac12cH_0.\eqno(36)$$ Hence, in the considered approach the
constant acceleration $w$ intrinsic to non-inertial RF $S'$ can
naturally be connected to the Hubble constant $H_0$, defining
space expansion.

\section{Time inhomogeneity and the blue-shifted frequency drift.
The Pioneer anomaly}

Now let us consider, on the basis of the formula (22), the
location-type experiments. The conventional scheme of such an
experiment is as follows:
\begin{itemize}
\item[(1)] the signal is emitted from the point of the observer
location at the time instant $t^0_A$,
\item[(2)] the signal is arrived and reemitted at the time instant $t_B$,
\item[(3)] the signal is returned to the observer at the time instant $t_A$.
\end{itemize}

Under the assumption of the coincidence of the forward
$(t_B-t_A^0)$ and backward $(t_A-t_B)$ time intervals, one can
obtain the formula for the signal traveling time
$t=\frac12(t_A-t_A^0)$, and then accept the formula $R=ct$ for the
corresponding location distance.

The time inhomogeneity (22) changes situation such that the
forward and backward time intervals do not coincide. The latter
time interval is larger then the former one (Fig. 4).

\begin{figure}[h!]
\centering \includegraphics[width=0.48\textwidth]{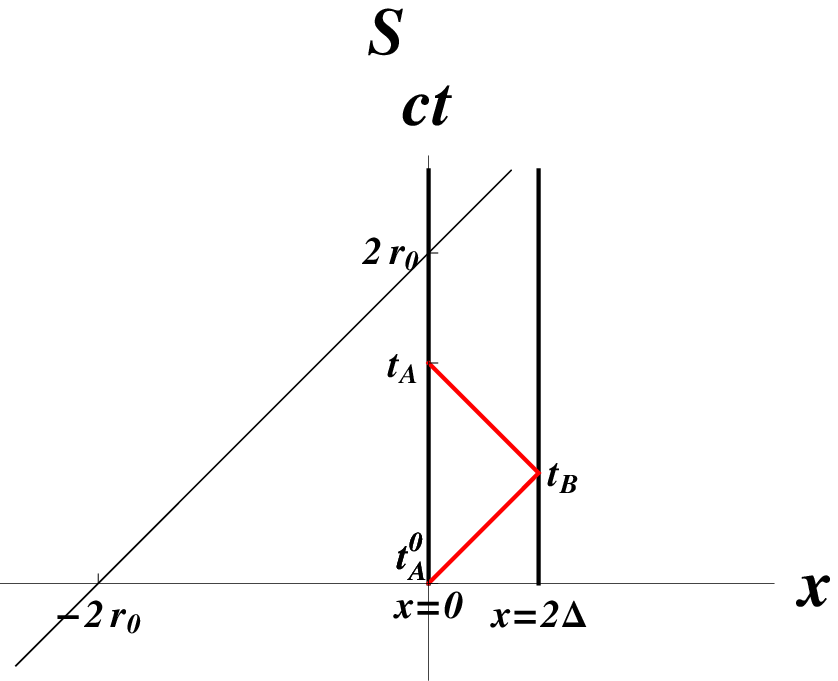}
\includegraphics[width=0.48\textwidth]{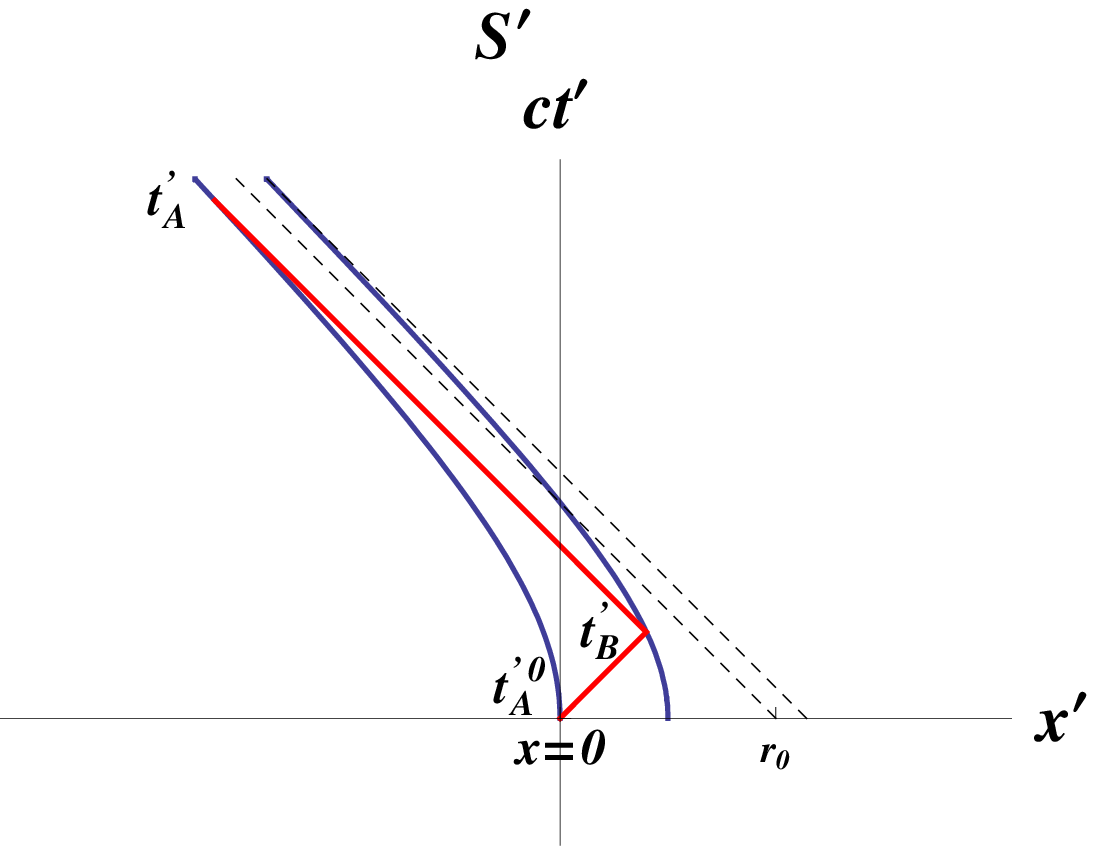} 
\caption{Location-type experiment. World lines and dashed lines
are as on Fig. 2. Thin lines are light signal world lines. Because
of conformal transformations, light lines remain straight and
inclined at $45^\circ$. In RF $S$ we have $t_A-t_B=t_B-t_A^0$, but
in $S'$ always the inequality $t'_A-t'_B>t'_B-{t'}_{\!A}^0$ takes
place.}
\end{figure}

In application to real experiment analysis one should use (22) for
the small time intervals, i.e., when $\Delta t/t_{lim} =
(t_A-t_A^0)/t_{lim} \ll 1$. In this case, formula (22) gives, to
the second order of $t/t_{lim}$ $$t'_\pm=t\mp\frac{t^2}{t_{lim}}.
\eqno(37)$$ In accordance with the location distance definition,
the signal travelling distances in forward and backward
directions, respectively, will be
$$x'_\pm=ct'_\pm=x\mp\Delta x=x\mp \frac{W_0}{2}t^2, \eqno(38)$$
where $$W_0=\frac{2c}{t_{lim}}=2w.\eqno(39)$$

Therefore in the $t/t_{lim} \ll 1$ approximation ($t$ is the
signal propagation time) the forward and backward location
distances $x'_\pm$ differ from $R=ct$ by $\Delta
x=\frac{W_0}{2}t^2$. From the usual point of view, it appears that
the emitter fixed in any space point is subjected to constant
acceleration $W_0=2w$ directed to the observer.

Clearly, $t/t_{lim} \ll 1$ condition is equivalent to the
condition $\frac{wt}{2c}\ll1$, $\frac{wx}{2c^2}\ll1$ (see Sec. 1),
which formally corresponds to Galilei-Newton kinematics. So, the
predicted effect in the location-type experiments will be the blue
frequency shift, which value will be defined by the formula
similar to (17), i.e.
$$\Delta \nu_{obs}=\nu_0\frac{W_0t}{c}.\eqno(40)$$ Due to (39) the value
of this shift is twice as large as predicted by (17).

For the constant rate of frequency shift
$\dot\nu_{obs}=\frac{d(\Delta \nu_{obs})}{dt}$, we find the
following relation analogous to (18):
$$\dot\nu_{obs}=\frac1c\nu_0W_0. \eqno(41)$$ Since by (35), $W_0=cH_0$,
where $H_0$ is the Hubble constant, from (41) we have
$$\dot\nu_{obs}=\nu_0H_0. \eqno(42)$$ This relation defines the
frequency drift as a function of the fixed emitter frequency
$\nu_0$.


According to the approach under consideration the anomalous
blue-shifted drift is the consequence of the background
acceleration existence (i.e. non-inertiality of the observers RF).
It can be observed in principle under suitable conditions (in the
absence of any gravitating sources) on any frequency even in the
case of mutually fixed emitter and receiver. From this point of
view PA should be treated as the first clearly observed effect of
that kind. The experimental data of Pioneer tracking can be used
for determination (or at least for estimation) of the parameter
$t_{lim}$ and corresponding background acceleration.

On the other hand, the uniform blue-shifted drift
$(\dot\nu_{obs})_P$ is measured experimentally with a great
accuracy $(\dot\nu_{obs})_P=(5.99\pm0.01)\cdot10^{-9}$ Hz/s
[1--4]. Therefore it can form a basis for the new (alternative to
the cosmological observations) high precision experimental
estimation of the numerical value of the Hubble constant.

For that goal, we make use of (42), and recall that frequency of
Pioneer tracking is $(\nu_0)_P=2.29\cdot10^{9}$ Hz, such that
$$H_0=\frac{(\dot\nu_{obs})_P}{(\nu_{obs})_P}
\cong2.62\cdot10^{-18}\ \mbox{s}^{-1},\eqno(43)$$ what is
consistent with generally accepted value of
$H_0\cong2.4\cdot10^{-18}\ \mbox{s}^{-1}$ obtained from
cosmological observations.

For the ''acceleration'' $a_P$ we have
$$a_P=cH_0=7.85\cdot10^{-8}\ \mbox{cm/s}^2,$$ what is in the range
of uncertainty of PA data ($a_P^{exp}=(8.74\pm1.33)\cdot10^{-8}\
\frac{\mbox{cm}}{\mbox{s}^2}$).

The numerical coincidence of the results can be considered as
experimental evidence of anomalous blue-shifted drift as a
kinematical manifestation of the conformal time inhomogeneity. In
other words, from the view point of the considered approach, the
quantities measured in experiments of electromagnetic wave
propagation favor relation (40) (but not (17)) for the anomalous
frequency shift.

It should be stressed that the physical meaning of the relations
(17) and (40) is fundamentally different in spite of their visual
similarity.

Equation (17), defining blue frequency shift by the
non-inertiality of RF $S'$, in fact was obtained in the
Galilei-Newton kinematics. There time transformation under
transition from RF $S$ to $S'$ has the form $t=t'$ (see (4)), and
the velocity transformations (16) as well as the acceleration $w$
are defined as usually in the Galilei-Newton kinematics.

On the other side, formula (40) was obtained from the exact
non-linear time transformation (22) defining the time
inhomogeneity, that is, beyond the Galilei-Newton kinematics.
``Constant acceleration'' $W_0=2w$ appears due to the quadratic
character of the first non-linear term in the power series
expansion of $t'(t)$ in terms of small parameter $t/t_{lim}$ in
(22), while the location distance is defined as $R'=ct'$.

Hence the ``acceleration'' $W_0$ is not a ``truly'' acceleration
(i.e. its origin is not a force or a dynamical source) but rather
a ``mimic'' acceleration. This imitation appears because of the
effects originated from non-linearity of time course (time
inhomogeneity), and interpreted in terms of traditional paradigm
of time homogeneity.

The possibility of assigning the anomalous frequency drift
observed in the signals transmitted by Pioneer 10/11 to the
quadratic time inhomogeneity was noted in the first comprehensive
works on PA [1] and [2]. However no theoretical reasons were
mentioned for such a phenomenological approach.

Concerning the acceleration parameter $w=\frac12cH_0$ we note that
it seems to be a strictly natural candidate for the ``minimal
acceleration'' of Milgrom, which is a fundamental dynamical
parameter of the Modified Newton Dynamics (MOND) (see [13, 14]),
which in turn is a phenomenological alternative for the Cold Dark
Matter approach. This question will be subject of a separate
publication.

\section{The possible kinematic origin of the accelerating Universe
phenomenon}

Now we analyze the general relation (32) for $V(z)/R(z)$
$$\frac{V(z)}{R(z)}=cR_u^{-1}\frac{(z+1)^{1/2}}{(z+1)^2+1}
\cdot\frac{(z+1)^2-1}{(z+1)^{1/2}-1}.$$ Taking into account the
connection (35) between $R_u$ and the Hubble constant $H_0$, we
rewrite this equation in the dimensionless form as
$$\phi(z)=\frac{V(z)}{H_0R(z)}= \frac12\frac{(z+1)^{1/2}}{(z+1)^2+1}
\cdot\frac{(z+1)^2-1}{(z+1)^{1/2}-1}.\eqno(44)$$

The function $\phi(z)$ and its derivative $\phi'(z) = \frac{d
\phi(z)}{dz}$ are shown in Fig. 5 and Fig. 6. Horizontal line in
Fig.~5 represents strict Hubble law (34). We see that this
function possesses a maximum at $z_0\cong0.475$ (see Fig. 5 and
Fig. 6). Overall variation of $\phi(z)$ demonstrates that in the
interval $0\leq z<z_0$ the distance $R(z)$ increases with $z$ more
slowly, and in the interval $z_0<z<\infty$ approaches its limit
value $R_u$ more rapidly, than the velocity $V(z)$ approaches its
limit $c$ (Fig. 7).




\begin{figure}
 \begin{center}
 \includegraphics[scale=0.8]{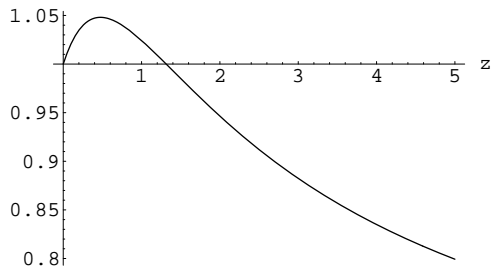}
 \caption{The function $\phi(z)$. }
 \label{Figure5}
 \end{center}
 \end{figure}

 \begin{figure}
 \begin{center}
 \includegraphics[width=0.48\textwidth]{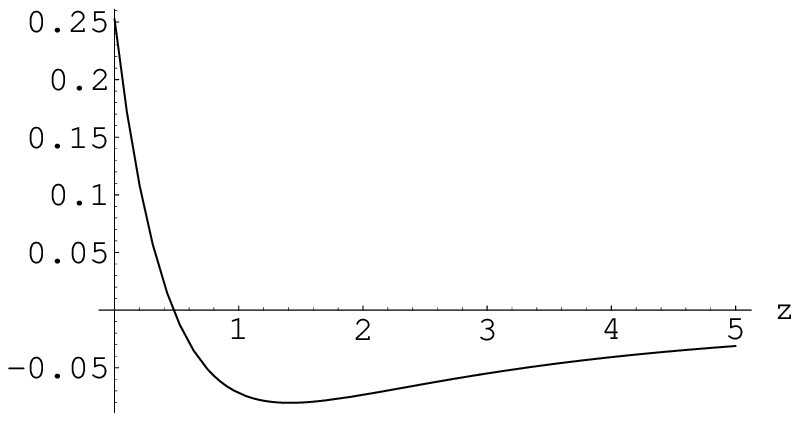}
 \includegraphics[width=0.48\textwidth]{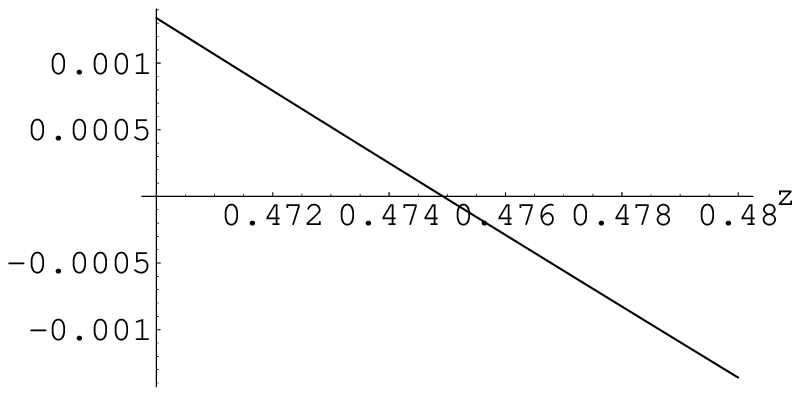}
 \caption{The derivative $\phi'(z)$ and position of its zero. }
 \label{Figure6}
 \end{center}
 \end{figure}

 \begin{figure}
 \begin{center}
 \includegraphics[scale=0.8]{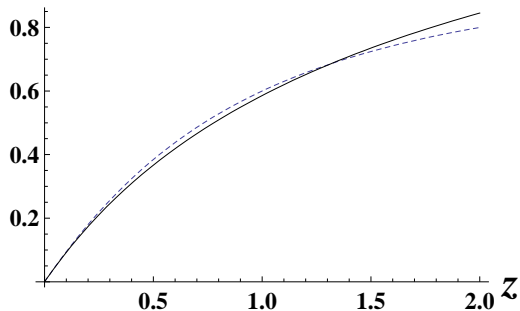}
 \caption{Functions $V(z)/c$ (solid line) and $R(z)/R_u$ (dashed
 line). The values of the both functions coincide at the points of
 $z=0$ and $z\approx1.315$.
 The inequality $V(z)/c>R(z)/R_u$
 ($V(z)/c<R(z)/R_u$) takes place in the interior (exterior) of this
 interval.}
 \label{Figure7}
 \end{center}
 \end{figure}

As regards a possible treatment of the behavior of the function
$\phi(z)$ in terms of the standard dynamical GR approach using the
decelaration parameter, we are to notice the following. According
to the pure kinematic approach proposed in our paper, the source
of the effects induced by the cosmologic expansion is the
conformal time inhomogeneity. The ``acceleration'' attributed to
the emitting source arises because of treating the actual
nonlinear time dependence $t'(t)$ in terms of the traditional
theoretical paradigm based on the time homogeneity concept. The
uniform ``acceleration'' $W_0=cH_0$, which appears in the formula
(38) in the $t/t_{lim}\ll 1$ approximation, is not a ``true'' but
the ``effective'' acceleration in reality. Such an
``acceleration'' in the general case must be treated as
time-dependent one. Purely formally, it can be determined as the
second time derivative of the function
$$
R(t)= c t'= c t\left(1-\frac{t}{t_{lim}}\right)^{-1} \eqno{(45)}
$$
and it takes the following form
$$
W(t)=\frac{2c}{t_{lim}}\left(1-\frac{t}{t_{lim}}\right)^{-3} .
\eqno{(46)}
$$
Such a ''time-dependent effective acceleration'' is directed to
the point of observation and coincides in the first order in
$t/t_{lim}$ with $W_0=cH_0$. This ''acceleration'' can be
presented, according to (26), in the form
$$
W(z)=W_0\left(1+z\right)^{3/2}.
$$
Thus, one can say that the numerical value of such an
''acceleration'' decreases during the process of the Universe
expansion, starting from the very large magnitude ($z\gg1$).

Evidently, the interpretation of $\phi(z)$ behavior from the point
of view of common treatment seems as follows. In the interval
$\infty > z > z_0$, there is \underline{deceleration}
$\left(\frac{d\phi}{dz}<0\right)$ of cosmological expansion, which
turns to \underline{acceleration}
$\left(\frac{d\phi}{dz}<0\right)$ at $z_0\cong 0{,}475$ it changes
to the. Numerical value of $z_0$ agrees quite well with
experimentally founded ''point of change'' $z_{exp}=0.46\pm0.13$.

It should be emphasized that the basic formula (22) for the
conformal transformations of the time, as well as all its
consequences, are valid on the assumption that the Hubble
parameter $H_0$ is constant. Hopefully this assumption is
reasonable as applied to at least later stages of the Universe
evolution. In this case the proposed formulae (28) and (32) can be
valid for the experimentally  obtained values of the read shift
having the order of several units.

\section{Summary and outlook}

The main results of the proposed approach consist in derivation of
kinematical consequences of the conformal deformation of the time,
which have not yet received the proper attention so far. The first
of these consequences is the possibility to express the
propagation time of the electromagnetic signal traveling from the
remoting source to the point of observation as an explicit
function on the red shift. That gives the possibility to obtain a
simple general expression defining dependence of the cosmological
distance on the red shift.

It would be of interest to correlate the obtained formula with the
cosmological observation data in the range of the red shift values
accessible experimentally. We stress that the obtained expression
for the cosmological distance dependence on the red shift follows
directly from the conformal transformations. Likewise, the well
known relativistic formula for the source --- receiver relative
velocity dependence on the frequency shift (Doppler effect)
follows from the Lorenz transformations in a pure kinematical way.
In doing so, the light-cone equation is used as an additional
condition in both cases.

The proposed derivation of the expression for the speed of
recession/intergalactic distance ratio is therefore a proof that a
Hubble law can be derived as a direct kinematical outcome of the
conformal group transformations.

The obtained expression can be verified experimentally, at least
in the accessible range of the red shift. It is remarkable that
the correspondence function in the red shift possesses a maximum
at the point $z_0\cong0.475$. This numerical value corresponds
surprisingly well to the experimentally registered value
$z_{exp}=0.46\pm0.13$.

The proof of existence of the direct connection between the
acceleration parameter of the conformal transformations and the
Hubble constant is the all-important conclusion of the proposed
derivation of the generalized expression for the Hubble law. The
existence of such a connection gives the theoretical foundation
for treatment of the observable local effects of the cosmological
expansion as a manifestation of the non-inertial (non-Lorentz)
character of the observer RF.

According to the proposed treatment such an effect can be detected
in the experiments dealing with signal propagation along the
closed path. The location-type experiment is the most simple one
from viewpoint of the possibility of its adequate description in
the frame of the spatially one-dimensional model.

The parameter $t_{lim}$ defining the upper limit of the duration
of signal propagation coincides by an order of magnitude with the
age of the Universe. It is much longer than the duration of any
really feasible location-type experiment, therefore it is
sufficient to restrict ourself with the first order approximation
in $t/t_{lim}\sim H_0t$ in the general formula for the conformal
time deformation. In such an approximation, the time inhomogeneity
becomes quadratic, and, consequently, the corresponding blue shift
gets the linear dependence on time. As it is known, the
experimentally observed PA-effect presents precisely such a shift.

In the framework of the conventional consideration assuming time
homogeneity, such a dependence allows for the only possible
interpretation as some additional uniform acceleration experienced
by signal source and directed towards observer.

According to the proposed approach, this background acceleration
is responsible for the PA which is manifestation of the
non-inertiality of the observer RF caused by acceleration. One can
say that the PA-effect has revealed the non-inertial character of
the ''expanding'' RF in the way analogous to the well-known Sagnac
effect which is a direct observed manifestation of the
non-inertiality of the rotating RF.

By the proposed interpretation of PA, the measured frequency shift
can be considered as a new independent high-precision measurement
of the Hubble constant. Moreover, the concept as a whole can be
directly tested in the experiment. The use of the appropriate
sources and monochromatic detectors at required experimental
conditions enables observation of the described anomalous blue
shift in its "pure form", when the source and detector are
mutually motionless, practically at all frequencies. As the effect
is linearly growing with the frequency, it is expedient to use
high-frequency radiation. In principle, this allows for a
considerable decrease in the observation time. To study this
effect, it is required to provide
\begin{itemize}
\item[(i)] maximum immobility of the source and detector,
\item[(ii)] elimination of the gravitational field effect of massive
bodies, \item[(iii)] minimization of thermal fluctuations and
mechanical deformations.
\end{itemize}
One should not rule out a possibility of providing all these
conditions in zero gravity at the satellites orbiting along the
circumterrestrial orbits. This proposal has been put forward in
[15].

As it was shown, the value of the mimic ''acceleration''
$W_0=cH_0$ defined from the experiments dealing with the
electromagnetic signal propagation, is double of the value of
relative acceleration $w$ playing the role of the conformal
transformation parameter. At the same time, the quantity
$w=\frac12cH_0$ coincides with the acceleration of any test
particle defined by the non-inertial observer in the absence of
any real force sources. This quantity can be correlated with the
''minimal acceleration'' in the MOND concept. One can notice that
in the distinct papers devoted to the MOND concept, different
numerical values of the Milgrom parameter appear, but in any case
they are defined as a fraction of $cH_0$.

All the conclusions of the present paper were obtained by the
assumption that the Hubble constant is time-independent. Needless
to say that such an assumption does not mean that the observable
magnitude of the Hubble constant is its lower limit. Any rigorous
theoretical arguments in favour of such an assumption are absent
today. Only some reasoning which are noting more than a suggestive
ones may be cited. They are connected with so-called Mass
Dependent Maximal Acceleration (MDMA) concept (see [16] and
references within) together with the Maximal Tension (or Maximal
Force) hypothesis firstly proposed in [17] and independently in
[16]. If we define, following Gibbons [17], the Maximal Force
$F_0$ as $F_0=\frac{c^4}{4G}$ ($G$ being the Newton gravitational
constant) the corresponding MDMA for the mass $M$ will be defined
as $W(M)=\frac{F_0}{M}$. Putting the Universe ''diameter'' equal
to $R_u=2cH_0^{-1}$ and defining the Universe ''mass'' $M_u$ as a
product of the critical density $\rho_c=\frac{3H_0^2}{8\pi G}$ and
the Universe ''volume'' $V_u=\frac{4\pi}{3}
\left(\frac{R_u}{2}\right)^3$ we obtain $W(M_u)=\frac12cH_0$ i.e.
the value coinciding with the background acceleration magnitude.

It is evident that such an intriguing coincidence is to be
confirmed in the framework of some consistent theoretical scheme.
The problem connected with the maximum tension hypothesis as
applied to the cosmological problems call for special
considerations in further publications of the author.

It should be noted that the existence of the background
acceleration resulting from the Methagalaxy expansion opens the
nontrivial alternative possibility to treat the dark matter
phenomenon (first of all - the dark energy one) as a peculiar kind
of the inertial forces manifestation. In this case the future of
the dark matter concept  would become similar to the destiny of
the substantional ether model abolished in due time by the Special
Relativity kinematics.

The proposed approach, in itself, is a certain natural (and
minimal) extension of Special Relativity concept, with evident
adherence to the correspondence principle (the availability of the
correct limiting transfer to the conventional physics when the
dimentionless parameter $tH_0$ trends to zero) .

In the present paper we restrict ourself to the use of spatially
one-dimensional model. But the results obtained are applicable to
the description of the effects associated with the longitudinal
components of the relative motion only. The location-type
experiments and the observation of signals from the cosmological
distant objects (i.e. from the past) are experiments of that kind.
The most simple line of taking into account the uniformity and
isotropy of the cosmological expansion in the framework of the
proposed approach is to choose the four-vector-parameter $a^\mu$
in the following form $a^\mu=(0,-\frac{w}{2c^2}\frac{\vec r}{r})$.
The acceleration $\vec W=w\frac{\vec r}{r}$, of course, unlike the
one-dimensional case, cannot be connected with either of known
physical forces. Nevertheless such a form of parameter $a^\mu$
seems to be quite acceptable for the phenomenological description
of the cosmological expansion. This question is now under
investigation.

It goes without saying that the proposed approach must be driven
in conformity with a General Relativity concept, in any case, in
the part of the GR using the conventional model of inertial
(Lorentz) observer. The matter is that the existence of nonzero
background acceleration caused by Universe expansion means
non-adequateness of such a model to physical reality. Really any
observer carrying out local measurements at any point of Universe
can detect background acceleration existence. As such he comes to
an unavoidable alternative: either his RF is non-inertial, or the
RF is inertial but there is some external constant gravitational
field. In any case the observer is not a Lorentz one.

It seems to be possible that the needed modification of
traditional gravitation theory can be accomplished by replacement
of the geometry of tangential space from Poincare to conformal
one.

\section{Acknowledgments}

The author would like to acknowledge E.V. Doktorov, V.V.
Kudryashov, A.E. Shalyt-Margolin,  Ya.M. Shnir, I.A. Siutsou  and
Yu.P. Vybly for their invaluable critical remarks and for the
fruitful discussions. The paper is in part supported by the
Belarus Foundation for Fundamental Research (BFFR), Project
F06-129.



\begin{thebibliography}{99}
\bibitem{1} J.D. Anderson et al., Phys. Rev. Lett. {\bf 81}, 2858--2861 (1998);  arXiv:gr-qc/9808081.
\bibitem{2} J.D. Anderson et al., Phys. Rev. D. {\bf 65}, 082004/1--50 (2002);  arXiv:gr-qc/0104064.
\bibitem{3} S.G. Turyshev, M.M. Nieto, and J.D. Anderson, arXiv:gr-qc/0503021.
\bibitem{4} S.G. Turyshev, M.M. Nieto, and J.D. Anderson, arXiv:gr-qc/0510081.
\bibitem{5} M. Mizony and M. Lachieze-Rey, arXiv:gr-qc/0412084.
\bibitem{6} A.O. Barut and R. Raczka. \textit{Theory of Group Representations and Applications}
(PWN - Polish Scientific Publishers, Warszawa, 1977).
\bibitem{7} L.M. Tomilchik, arXiv:gr-qc/0704.2745.
\bibitem{8} L. Page, Phys. Rev. {\bf 49}, 254, 946 (1936); L. Page and
N.I. Adams, ibid. {\bf 49}, 466 (1936).
\bibitem{9} T. Fulton, F. Rohrlich, and L. Witten, Rev. Mod. Phys. {\bf 34}, 442 (1962).
\bibitem{10} T. Fulton, F. Rohrlich, and L. Witten, Nuovo Cim. {\bf 34}, 652 (1962).
\bibitem{11} R.L. Ingraham, Nuovo Cim. {\bf 12}, 825 (1954).
\bibitem{12} W.L. Burke, \textit{Spacetime, Geometry, Cosmology} (University Science
Books, Mill Valley, California, 1980).
\bibitem{13} M. Milgrom, Astrophys. J. \textbf{270}, 365 (1983).
\bibitem{14} J. Bekenstein and M. Milgrom, Astrophys. J. \textbf{286}, 7 (1984).
\bibitem{15} L.M. Tomilchik, Optics Spectrosc. {\bf 103}, 237 (2007).
\bibitem{16} L.M. Tomilchik and V.V. Kudryashov, Proc. Int. School-Sem. "Actual
Problems of Microworld Physics", V 1, JINR, Dubna (2004);
arXiv:gr-qc/0507090.
\bibitem{17} G.W. Gibbons, Found. Phys. \textbf{32}, 1891 (2002); arXiv:hep-th/0210109.
\end{thebibliography}
\end{document}